\documentclass[prl,aps,epsfig,onecolumn]{revtex4}
\usepackage{epsfig,graphics}

\begin{document}

\title{Current-Induced Motion of Narrow Domain Walls and Dissipation in
Ferromagnetic Metals}

\author{M. Benakli}

\email{mourad.benakli@seagate.com}

\author{J. Hohlfeld}

\email{julius.hohlfeld@seagate.com}

\author{A. Rebei}

\email{arebei@mailaps.org}

\affiliation{\textit{Seagate Research Center, Pittsburgh, Pennsylvania 15222,
USA}}

\begin{abstract}
Spin transport equations in a non-homogeneous ferromagnet are derived
in the limit where the sd exchange coupling between the electrons
in the conduction band and those in the d band is dominant. It is
shown that spin diffusion in ferromagnets assumes a tensor form. The
diagonal terms are renormalized with respect to that in normal metals
and enhances the dissipation in the magnetic system while the off-diagonal
terms renormalize the precessional frequency of the conduction electrons
and enhances the non-adiabatic spin torque. To demonstrate the new
physics in our theory, we show that self-consistent solutions of the
spin diffusion equations and the Landau-Lifshitz equations in the
presence of a current lead to a an increase in the terminal velocity
of a domain wall which becomes strongly dependent on its width. We
also provide a simplified equation that predicts damping due to the
conduction electrons. 
\end{abstract}

\pacs{72.25.Ba,72.25.Pn, 75.40.Gb,75.40.Mg}

\date{\today}

\maketitle
Dynamics of magnetic domain walls (DW) is a classic topic \cite{landau,thiele,asselin}
that recently received a lot of attention due to new fabrication and
characterization techniques that permit their study at the nanometer
scale. Moreover, the subject of spin dynamics in the presence of large
inhomogeneities is currently of great interest experimentally and
theoretically due to the potential applications in various nano-devices,
especially magnetic storage \cite{parkinIBM}. One particular area
that is still not well understood is the interaction of domain walls
(DWs) with polarized currents. The question here is how best to represent
the contribution of the spin torque to the dynamics of the magnetization
\cite{yamaguchi,hayashi,beach,thiaville,zhang,tatara,simanek,barnes,berger,duine,rebei}
. So far attention has been focused on wide DWs where it was shown
that terminal velocities are independent of the DW width \cite{zhang,duine}.

This paper extends previous treatments to the case of thin, less than
$100$ nm, DWs. One of the main objectives of our work is to expose
the interplay between linear momentum relaxation and spin relaxation
as the conduction electrons traverse a thin DW. This interplay originates
from the strong exchange interaction between the conduction s electrons
and the localized d moments, and makes the terminal velocities as
well as the transport parameters of the conduction electrons dependent
on the configuration of the local magnetization. This leads to an
enhancement of the non-adiabatic contribution of the spin torque to
the DW motion and opens the way to study spin torque-induced magnetization
dynamics in thin DWs in greater depth by measurement of DW velocities.
Moreover, we show that the interaction of the conduction electrons
and the d moments is also relevant for homogeneously magnetized metallic
systems, where it is at the origin of intrinsic damping. Our work
can be easily adapted to magnetic multilayer structures and hence
the equations derived here are capable to treat non-collinear magnetization
geometries as opposed to that in ref. \cite{valet} which deal only
with collinear configurations. Narrow DWs can exist either naturally
\cite{tan,kubetzka} or artificially \cite{khizroev,aziz} and we
hope the results discussed here show the potential benefits of studying
dissipation in DW-like structures.

To derive the spin coupling of the s electrons to the magnetization,
we adopt the sd picture which has been the basis for most of the studies
in DW motion \cite{zhang}. In the following we use $(l,m,n)$ for
moment indexes, and $(i,j,k)$ for space indexes. In addition the
transverse domain wall is assumed to extend in the $x$ direction,
with magnetization pointing in the $z$ direction. We start from the
Boltzmann equation satisfied by the $2\times2$ distribution function
of the conduction electrons, $\mathbf{f}=f^{e}+\mathbf{f^{s}}\cdot{\bf \sigma}$,
where $\sigma_{l}\,(l=1,2,3)$ are Pauli matrices, in the presence
of the magnetization $\mathbf{M}$ of the system and an external electric
field $\mathbf{E}$: \begin{eqnarray}
\partial_{t}\mathbf{f}+{\bf v}\cdot\nabla\mathbf{f} &  & +e\left({\bf E}+{\bf v}\times{\bf H}\right)\cdot\nabla_{p}\mathbf{f}+\nonumber \\
i\left[\mu_{B}{\bf \sigma}\cdot{\bf H}_{sd},\mathbf{f}\right] &  & =-\frac{f^{e}-f_{0}^{e}}{\tau_{p}}-\frac{\mathbf{f}-\mathbf{f}_{0}^{s}}{\tau_{sf}}.\end{eqnarray}
 The sd exchange field is ${\bf H}_{sd}(\mathbf{x},t)=J\,{\bf M}(\mathbf{x},t)/\mu_{B}$
with $J\approx1.0$ eV, and $\tau_{p}$, $\tau_{sf}$ are the momentum
and spin relaxation times, respectively \cite{rebei,hirst,kaplan}.
The variables $\mathbf{v}$, $e$, and $\mu_{B}$ are the velocity,
the charge and the magnetic moment of the s electrons, respectively.
$f_{e}^{0}$ and $\mathbf{f}_{s}^{0}$ are the equilibrium charge
and spin distribution. 

The conduction electrons have a polarization $\mathbf{m}=\mu_{B}\int\frac{d{\bf p}}{\left(2\pi\right)^{3}}Tr{\bf \sigma}{\bf \mathbf{f}}$
and carry a charge current $\mathbf{j}_{c}=e\int\frac{d{\bf p}}{\left(2\pi\right)^{3}}{\bf v}Tr{\bf \mathbf{f}}$,
as well as a spin current \begin{equation}
{\bf j}_{s}=\int\frac{d{\bf p}}{\left(2\pi\right)^{3}}{\bf v}Tr{\bf \sigma}{\bf \mathbf{f}.}\end{equation}

In the following we use normalized definitions of the moments, i.e.
$||\mathbf{M}||=||\mathbf{m}||=1$. The d electrons will be assumed
to satisfy a Landau-Lifshitz-Gilbert (LLG) equation 
\begin{equation}
\frac{d{\bf M}}{dt}=-{\bf M}\times\left(\gamma{\bf H}_{eff}+\frac{1}{\tau_{ex}}{\bf m}\right)+\alpha_{{\rm pd}}{\bf M}\times\frac{d{\bf M}}{dt},\end{equation}
 where $\tau_{ex}$ is the inverse of the precessional frequency,
$\omega_{c}=J/\hbar$, of the conduction electrons due to the exchange
field. $\mathbf{H}_{eff}$ is the total field acting on the magnetization
which includes the exchange field between the d-moments, the demagnetization
field and the anisotropy field. In metals, the main source of dissipation
is believed to be due to the conduction electrons which in our theory
is accounted for explicitly within the limitations of the sd model
\cite{rebeiH}. Hence, the damping constant $\alpha_{{\rm pd}}$ is
assumed to be due to dissipation caused by channels other than the
conduction electrons such as phonons or defects.

In inhomogeneous magnetic media, the sd exchange term becomes comparable
to that of the Weiss molecular field and hence the effect of the conduction
electrons on the magnetization should be taken beyond the linear response
approach. Going beyond the linear theory will allow us to see how
the presence of the background magnetization affects the transport
properties of the conduction electrons. We believe this is especially
true in transition metal nano-magnetic devices where the hybridization
of the s and d electrons is strong. Using standard many-body methods
\cite{rebei}, the diffusion contribution to the spin current can
be found \begin{equation}
j_{s}^{li}\left(t,{\bf x}\right)=-{\cal D}^{ln}\left(t,{\bf x}\right)\nabla_{i}m_{n}\left(t,{\bf x}\right),\end{equation}
 where $\mathcal{D}$ is a diffusion tensor with effective relaxation
time $\tau$ which will be assumed equal to the momentum relaxation
$(\tau\approx\tau_{p})$. The $\mathcal{D}$ tensor obeys the reduced
symmetry of the ferromagnetic state and is \cite{rebei} \begin{equation}
{\cal D}=D_{\perp}\left[\begin{array}{ccc}
1+\tau^{2}\Omega_{x}^{2} & \tau\Omega_{z}+\tau^{2}\Omega_{x}\Omega_{y} & -\tau\Omega_{y}+\tau^{2}\Omega_{z}\Omega_{x}\\
-\tau\Omega_{z}+\tau^{2}\Omega_{y}\Omega_{x} & 1+\tau^{2}\Omega_{y}^{2} & \tau\Omega_{x}+\tau^{2}\Omega_{y}\Omega_{z}\\
\tau\Omega_{y}+\tau^{2}\Omega_{z}\Omega_{x} & -\tau\Omega_{x}+\tau^{2}\Omega_{y}\Omega_{z} & 1+\tau^{2}\Omega_{z}^{2}\end{array}\right],\end{equation}
 where $\Omega=J\mathbf{M}/\hbar$, $D_{\perp}=\frac{D_{0}}{1+\left(\tau\omega_{c}\right)^{2}}$
with $D_{0}={\frac{1}{3}}{\rm v_{f}}^{2}\tau_{p}$ being the diffusion
constant of the electron gas with Fermi velocity ${\rm v_{f}}$. It
should be observed that in the presence of spin-orbit coupling, the
symmetry of the diffusion tensor will be the same as given here but
the separation of the relaxation times in independent channels of
momentum and spin relaxation will not be valid. In the following,
the effect of the electric field is taken only to first order.

\ The symmetry of the spin current is best revealed by going to a
local frame where the magnetization lies in the z-direction. In this
frame, one obtains for $\mathbf{E}=\mathbf{0}$ \begin{equation}
j_{\perp}=-D_{{\rm eff}}\frac{d\mathfrak{m}}{dx},\;\; j_{z}=-D_{0}\frac{dm_{z}}{dx},\end{equation}
 where $\mathfrak{m}\left(x\right)=m_{x}\left(x\right)-im_{y}\left(x\right)$,
and $D_{{\rm {eff}}}=D_{\perp}+iD_{xy}$ is an effective diffusion
coefficient with $D_{xy}=D_{\perp}\tau\omega_{c}$. \ From the divergence
of the spin current we get the steady-state equation for the spin
accumulation, \begin{equation}
\frac{d^{2}\mathfrak{m}}{dx^{2}}=\frac{\mathfrak{m}}{\lambda_{{\rm eff}}^{2}},\;\;\frac{d^{2}m_{z}}{dx^{2}}=\frac{m_{z}-m_{0}}{\lambda_{sdl}^{2}},\end{equation}
 where $\lambda_{{\rm eff}}^{2}=\tau_{{\rm eff}}D_{{\rm eff}}$ with
$\tau_{{\rm eff}}=1/(\frac{1}{\tau_{sf}}-i\omega_{c})$, $m_{0}$
is the equilibrium spin density, and $\lambda_{sdl}$ is the longitudinal
spin diffusion length typically in the range of 5-100 nm.\ The general
solutions for the complex accumulation are of the form $\mathfrak{m}\left(x\right)=A\exp\left[-x/\lambda_{{\rm eff}}\right]+B\exp[x/\lambda_{{\rm eff}}]$,
i.e. they show an exponential decrease (or increase) and oscillations
from a local inhomogeneity in $\mathbf{M}$. \ In the limit of a
large sd exchange field the period of the oscillations is $\frac{{\rm v_{f}}}{\omega_{c}}$
which corresponds to the coherence length $1/|k^{\uparrow}-k^{\downarrow}|$
in the ballistic approach, where $k^{\uparrow}$ is the spin-up momentum.

Our expressions for the spin current generalize those used currently
in the literature \cite{zhang}. 
We find that the diffusion constant $D_{0}$ is now renormalized by
$1/(1+(\tau\omega_{c})^{2})$ which means that precession in the exchange
field reduces diffusion. Moreover, the precession gives rise to off-diagonal
terms in the diffusion tensor which reflect the local 2D rotational
symmetry around $\mathbf{M}$.

The origin of the off-diagonal term $\mathcal{D}_{xy}$ can be understood
qualitatively in terms of flux. First we rewrite it in the following
form \begin{eqnarray}
D_{xy}=(\frac{1}{3}{\rm v_{f}}^{2}\tau_{p})\frac{\tau_{p}\omega_{c}}{1+(\tau_{p}\omega_{c})^{2}}=\frac{1}{3}\frac{{\rm v_{f}}^{2}\omega_{c}}{\nu^{2}+\omega_{c}^{2}}\end{eqnarray}
 where $\nu=\frac{1}{\tau_{p}}$. In the limit of fast precession,
$\nu\ll\omega_{c}$, we have $D_{xy}=\frac{1}{3}{\rm v_{f}}^{2}/\omega_{c}$.
Next if we set ${{\rm v_{f}}/\omega_{c}=L_{m}}$, then $L_{m}$ is
the distance a spin typically goes before it 'converts' into the spin
at 90-degrees to that which it started with. The corresponding contribution
to the flux has an obvious interpretation - the source of spin x,
$m_{x}$, is particles coming from a distance $L_{m}$ away where
they had spin y, $m_{y}$. The flux can be derived from a simple `kinetic'
argument. A distance $L_{m}$ upstream, the density is $m_{y}=m_{y}^{0}+L_{m}\frac{dm_{y}}{dx}$
and a distance $L_{m}$ downstream, $m_{y}=m_{y}^{0}-L_{m}\frac{dm_{y}}{dx}$.
The flux of particles with spin x, $m_{x}$, crossing a point, coming
from upstream, is $m_{y}{^{\uparrow}}{{\rm v_{f}}/3}$ and from downstream
it is $m_{y}{^{\downarrow}}{\rm v_{f}/3}$. The difference is then
$(m_{y}{^{\uparrow}}-m_{y}{^{\downarrow}}){\rm v_{f}/3=2L_{m}\frac{dm_{y}}{dx}{\rm v_{f}/3}}$
which, within a factor of 2, is our off-diagonal flux. In short, the
off-diagonal terms are the corrections induced by precession on the
diffusion process.

To get the effective equation for $\mathbf{M}$, we use equation 3
to express $\mathbf{m}$ in terms of the magnetization $\mathbf{M}$,
then use the implicit solution back into the equation for $\mathbf{m}$,
Eq. 1. We find that the equation of motion for the magnetization becomes
\begin{eqnarray}
\beta\frac{d{\bf M}}{dt} & = & -\gamma{\bf M}\times{\bf H}_{eff}+\mathbf{a}\cdot\nabla{\bf M}+\left(\alpha_{{\rm pd}}-\xi\right){\bf M}\times\frac{d{\bf M}}{dt}\label{a1}\\
 &  & -\xi\gamma{\bf M}\times\left({\bf M}\times{\bf H}_{eff}\right)+\nabla\cdot{\cal D}\nabla{\bf m},\nonumber \end{eqnarray}
 where $\mathbf{a}=P(\mu_{B}/e)Tr\mathbf{j}$, $\beta=1+m_{0}+\alpha_{{\rm pd}}\xi$,
(or MB$\beta'=(1+m_{0}+\xi^{2})$ ), $\xi=\tau_{ex}/\tau_{sf}$ is
the ratio of the precessional time to the spin relaxation time of
the conduction electrons, and $P$ is the spin current polarization
\cite{zhang}. The second term on the right is the adiabatic spin
torque while the last term is the diffusion contribution. The third
and the fourth terms are equivalent to the non-adiabatic spin torque
with the original damping $\alpha_{{\rm pd}}$ from Eq.\,3 added
to the third term. For uniform magnetization, $\nabla\mathbf{M}=\nabla\mathbf{m}=0$,
and damping constant $\alpha_{{\rm pd}}=0$, Eq.\,\ref{a1} reduces
to \begin{eqnarray}
\frac{d\mathbf{M}}{dt} & = & -\gamma\frac{\xi^{2}+\beta}{\xi^{2}+\beta^{2}}\mathbf{M}\times\mathbf{H}_{{\rm eff}}\nonumber \\
 &  & -\gamma\xi\frac{m_{0}}{\beta^{2}+\xi^{2}}\mathbf{M}\times\left(\mathbf{M}\times\mathbf{H}_{{\rm eff}}\right).\label{b1}\end{eqnarray}
Hence, we are able to predict damping due to conduction electrons,
quantify the corresponding damping constant $\alpha_{el}=\gamma\xi m_{0}/(\beta^{2}+\xi^{2})$,
and identify its origin as the spin torque contribution of the conduction
electrons. We have written Eq. \ref{b1} in the LL form, but it equally
can be written in the LLG form. 
We have already shown in ref.\cite{rebei} that the magnetization
dynamics of a thin film embedded between two normal conductors and
subjected to an electric field is not well described by closed LL
(or LLG) equations.

Next we discuss qualitatively the effect of the diagonal and off-diagonal
terms of the diffusion tensor on the velocity of a domain wall, of
width $\lambda$. If we ignore the spatial dependence of the diffusion
tensor elements and replace the Laplacian in the diffusion equation
by $1/\lambda^{2}$, then we recover equations similar to those discussed
by Zhang and Li \cite{zhang} but with renormalized spin flip scattering
rate, $1/\tau_{sf}\rightarrow\,1/\tau_{sf}^{N}=1/\tau_{sf}+D_{0}/\lambda^{2}$,
and renormalized precessional frequency, $1/\tau_{ex}\rightarrow\,1/\tau_{ex}^{N}=1/\tau_{ex}-D_{xy}/\lambda^{2}$.
Therefore, the velocity and the effective damping of the DW dependent
on the size of the inhomogeneities in the magnetization. This can
be understood qualitatively from the results in \cite{zhang} which
showed that the DW velocity ${\rm v}$ for a wide DW, i.e.\,$\nabla\mathbf{m}\approx0$,
is inversely proportional to the damping $\alpha_{el}$ (in the case
$\alpha_{pd}$=0), ${\rm v}\approx(Pj\mu_{B}/e)((1+\xi^{2})/(\xi m_{0}))$. Then, ignoring the renormalization of the diffusion coefficient
$D_{0}$, the velocity is expected to take a similar form as in the
case which does not account for the diffusion but with $\xi$ replaced
by $\xi^{N}=\tau_{ex}^{N}/\tau_{sf}^{N}$. The damping $\alpha$ will
be also affected by this renormalization as is expected, since broadening
due to inhomogeneities is well known to occur in ferromagnetic resonance
measurements.

Now, we turn to the discussion of the results of the above theory
for a 1-D DW configuration. 
We solve numerically the coupled equations of motion for the conduction
electrons and that of the magnetization. We include the d-d exchange
between the local moments, the anisotropy along the direction of the
current and the dipole field. Pinning is neglected but can be easily
included in the simulations. Besides varying the width of the DW,
we also vary the other parameters in the sd model since there is no
universal agreement on their exact values. For example, it is generally
believed that spin relaxation times are about two orders of magnitude
longer than momentum relaxation times. While this may be true in paramagnets
, we already know that in Ni$_{80}$Fe$_{20}$ they are comparable
\cite{fert}. In Permalloy, the spin diffusion length, $l_{s}={\rm v_{f}\sqrt{\tau_{sf}\tau_{p}}}$,
is of the order of $5$ nm which is of the same order as the mean
free path, $l_{p}=3$ nm. 
%
\begin{figure}[ht]

\begin{centering}
\mbox{\epsfig{file=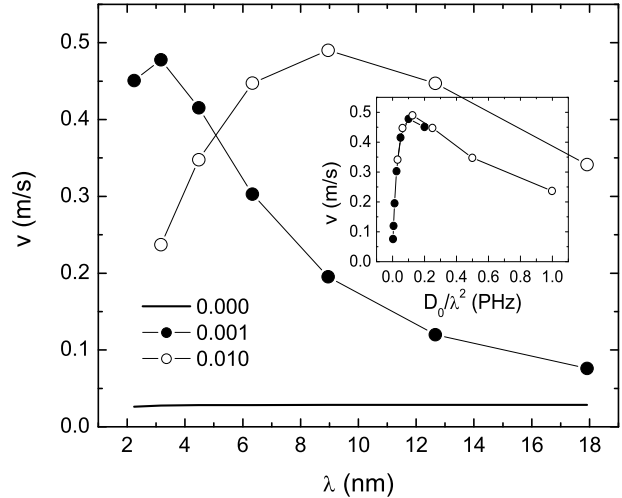,height=8 cm}} 
\par\end{centering}

\caption{\label{fig1} Domain wall velocity as a function of domain wall width
for ${\rm j_{c}=10^{8}}$ A/cm$^{2}$, $\tau_{sf}=3.0\times10^{-12}\,$s,
$\alpha_{{\rm pd}}=0.01$ and three different diffusion coefficients,
$\mathcal{D}=D_{0}\mathbf{I}$, given in the figure in units of m$^{2}$/s.
A value of $D_{0}=10^{-2}\,$m$^{2}$/s corresponds to $\tau_{p}\approx10^{-14}\,$s.
The solid thick line is that of Zhang and Li \cite{zhang}. The inset
shows the DW velocity versus $\frac{D_{0}}{\lambda^{2}}$.}
\end{figure}

Figure 1 shows the effect of introducing the (unnormalized) diffusion
term $D_{0}$ in the equations of motion of the magnetization. For
DW width larger than $100$ nm our result approximately recovers that of Ref. \cite{zhang}.
The variations of the domain wall velocity ${\rm v}$ with $\lambda$
are found to depend strongly on $D_{0}$. This is expected since ${\rm v}$
is, to first order, a function of $D_{0}/\lambda^{2}$ (cf. inset).
Moreover, the velocity peaks when the mean free path of the conduction
electrons, $l_{p}$, is of the same order as the DW width, since for
$l_{p}\gg\lambda$ there is almost no scattering while for $l_{p}\ll\lambda$
there is only slow diffusion. We have confined our results to $\lambda \ge 2$ nm since at much smaller 
DW widths, we expect contributions from Coulomb interactions and a breaking of the 
quasiclassical picture employed here.


In figure 2, we show the effect of the corrections introduced by the
off-diagonal terms in the diffusion tensor. This non-adiabatic effect
actually appears to suppress the DW velocity or the effect of diffusion
as we explained earlier. Otherwise, the functional behavior of the
velocity remains similar to the one discussed in figure 1. %
\begin{figure}[ht]

\begin{centering}
\mbox{\epsfig{file=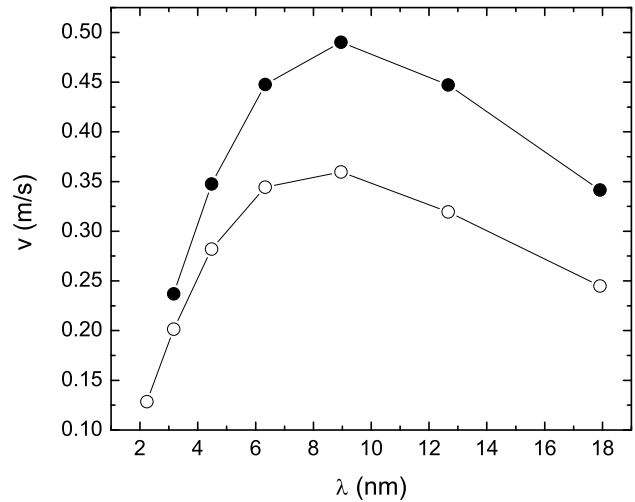,height=8 cm}} 
\par\end{centering}

\caption{\label{fig2} Domain wall velocity as a function of domain wall width
with the correct diffusion tensor taken into account. The solid (open)
symbols are without (with) off-diagonal corrections of the diffusion
tensor. Parameters are identical to those in Fig.\,\ref{fig1}}
\end{figure}

Finally in figure 3, we extract the contribution of the conduction
electrons to the effective damping of the magnetization. First, we
observe that the off-diagonal diffusion terms have little effect on
the relaxation of $\mathbf{M}$ which is mainly determined by the
spin relaxation time $\tau_{sf}$. These results are also not sensitive
to the DW width and the extracted electronic damping has the correct
order of magnitude for metals. 

%
\begin{figure}[ht]
\begin{centering}
\mbox{\epsfig{file=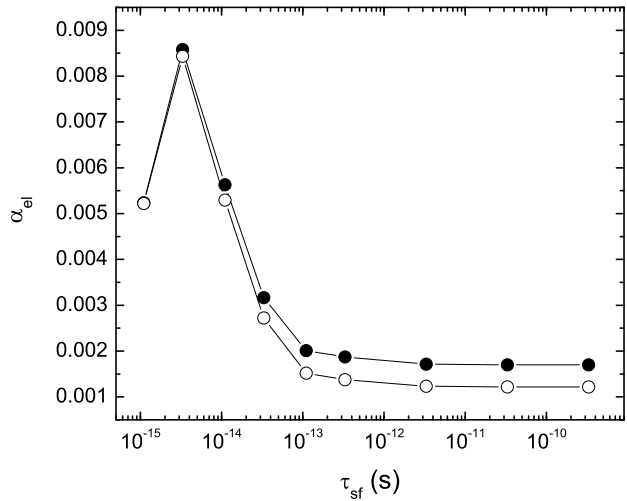,height=8 cm}} 

\par\end{centering}

\caption{The electronic damping $\alpha_{el}$ as a function of spin flip
scattering $\tau_{sf}$ for a $10\,$nm domain wall. The solid (open)
symbols are for off-diagonal terms included (not included). The diffusion
constant is $D_{0}=10^{-2}m^{2}/s$ and $\alpha_{{\rm pd}}=0$.}
\end{figure}

In summary, we have solved the conduction electron-magnetization problem
in the presence of a current self-consistently. We found that the
diffusion term provides a larger contribution to the drive torque
than to the damping process, leading to a overall increase of domain
wall velocity. We also showed that the new off-diagonal terms of the
diffusion tensor enhance the DW velocities which become at least one
order of magnitude larger than previously found. Moreover, the dependence of the
DW velocity on the width of the DW was found to be non-linear and
strongly dependent on the non-adiabatic behavior of the conduction
electrons through the non-diagonal corrections of the diffusion tensor.
We have been also able to determine the contribution of the conduction
electrons to the damping in ferromagnetic metals which we found to
be of the same order as the typical measured values of $\alpha$.
Therefore, our treatment allows us to include electronic damping in
micromagnetic calculations in a more rigorous way than is currently
done by simply accounting for it by a simple $\alpha$ parameter.


We thank W. N. G. Hitchon, E. Simanek, L. Berger, and P. Asselin for
related useful discussions.

\end{document}